\documentclass[10pt, conference]{IEEEtran}
\usepackage{graphicx}
\usepackage{amsmath}
\usepackage{textcomp}
\usepackage{amssymb}
\usepackage{bbm}
\usepackage{array}
\usepackage{fancyhdr}
\usepackage{amsthm}
\usepackage{wasysym}
\usepackage[normal]{subfigure}
\usepackage{pifont}
\usepackage[section]{placeins}
\usepackage{latexsym}
\usepackage{multirow}
\usepackage[dvips]{epsfig}
\DeclareMathOperator*{\argmin}{\arg\min}
\begin{document}
\title{Obstacle Constrained Total Area Coverage\\in Wireless Sensor Networks}
\author{
\authorblockN{Shyam~M}
\authorblockA{Department of Electrical Communication Engineering\\
Indian Institute of Science \\
Bangalore, 560012, India\\
Email: shyam@ece.iisc.ernet.in} \and
\authorblockN{Anurag~Kumar}
\authorblockA{Department of Electrical Communication Engineering\\
Indian Institute of Science \\
Bangalore, 560012, India\\
Email: anurag@ece.iisc.ernet.in}}
\maketitle
\begin{abstract}
This paper deals with the accomplishment of total area coverage of an arbitrary region using sensors with a finite sensing radius of $r_s$. For a given region, we aim to obtain a deterministic placement of sensors which, apart from ensuring that the entire region comes under the purview of at least a single sensor, minimises the number of sensors utilised. We begin by considering regions devoid of obstacles and thus having every location amenable for placement. Herein, we formalise the popular notion that sensors at the centres of the hexagons of a hexagonal tessellation provide the most optimal placement. We then move on to regions which may comprise obstacles of arbitrary size at arbitrary locations. We recognise two distinct classes of obstacles, namely transparent and opaque obstacles, which are distinguished by their ability (or the lack of it) to permit sensing radiation through them. In the real world, transparent obstacles model lakes, ponds and swamps, while the opaque ones stand for, inter alia, hills, trees and walls. We propose a polynomial-time algorithm for achieving optimal placement in the aforesaid scenarios and we prove its convergence. 
\end{abstract}
\begin{IEEEkeywords}Coverage, Transparent and opaque obstacles, Art gallery problem\end{IEEEkeywords}
\section{Introduction}
\label{sect: intro}

A wireless sensor network (WSN) consists of a large number of tiny, low-powered devices, called sensors, which communicate with 
each other (possibly) in a multi-hop fashion. These sensors suffer from severe constraints, such as low reliability and limited battery 
power (or energy), processing power, storage memory, sensing range and communication range capabilities, to name a few, with 
energy being the most critical one. Tremendous advances in inexpensive sensor technology and wireless communications have rendered the design and development of large-scale wireless sensor networks cost-effective and viable enough to attract the attention of a wide range of civilian, natural, and military applications.

One of the fundamental issues that arises in sensor networks, in addition to location calculation, tracking and deployment, is 
coverage. Coverage can be construed to be a measure of quality of service of a sensor network, in the sense that a high degree of coverage of a region signifies that the region is monitored well. Let us consider a specific application where coverage of a region becomes a critical necessity. Millions of acres of land are lost every year around the world due to forest fires. An early detection of these fires plays a crucial role in preventing them from causing irremediable calamity. It is possible to employ specialised sensor nodes in critical areas to monitor them and thereby preclude the occurrence of such disasters. The benefits of the placement of a large number of nodes to cover such areas clearly override the cost factor. It is fairly evident that coverage is a QOS parameter in this example, as the odds of a fire being detected early is proportional to the extent of coverage employed. 

The problem of area coverage is thus elementary in the design of wireless sensor networks. In this paper, we aspire to design algorithms that could achieve total area coverage in an arbitrary two-dimensional region. A two-dimensional region is said to be \emph{1-covered} if every point (excluding the obstacles) inside the region is within a distance of $r_s$ from its nearest sensor, where $r_s$ is the sensing radius of the sensor. Wireless sensor nodes (or motes) have to be placed in the region to obtain complete coverage. The sensor design and the sensing modality is such that the coverage region of a sensor is a disk of radius $r_s$ centred at the location where the sensor is placed. The region to be covered has \emph{transparent obstacles} (e.g., lakes, ponds and swamps) where the sensors cannot be placed but through which the signal being sensed can pass, as well as \emph{opaque obstacles} (e.g., trees and other structures), which neither allow the sensors to be placed nor allow the signal being sensed to penetrate through them. The problem we deal with is to cover the area not occupied by obstacles with as few as sensors as possible, ensuring that a specific coverage objective (e.g., $k$-coverage) is met. 
\section{Related Work}
\label{sect:litsurvey}
There are three main classes of coverage problems tackled in the literature on coverage. These are barrier coverage, area coverage and point coverage. Barrier coverage involves placing sensors along vulnerable boundaries of a region for achieving intrusion detection. Area coverage refers to the placement of sensors to monitor the entire area, while point coverage focusses on the monitoring of certain critical points identified within a region. We now cast a quick glance over some of the specific coverage problems dealt with in the literature. 

Gautam Das et al. \cite{das-06-BS_cover_Convex_area} provide a $O(k\log k)$ algorithm to evaluate the placement of an optimal number of sensors to cover a given region without obstacles. An initial random placement of the $k$ sensors is followed up with the Voronoi partition of the area. In every iteration of the algorithm, the position of the sensor is moved towards the circumcircle of the Voronoi polygon. It is proved that the new position of the sensor lies within the same polygon and also that the radius of the circumcircle reduces with every iteration. The procedure terminates when the radius of the circle does not shrink anymore.

The above algorithm can be adapted to solve the problem of optimal sensor placement in a region without obstacles. We can start with an arbitrary number of sensors and follow the procedure. An appropriate value of the number would be the ratio of the area of the region to the sensing area of a sensor. If the algorithm terminates with the maximum radius of all the circumcircles being greater than the typical sensing radius of a sensor, the iteration needs to restart with more number of sensors.

Senjuti Basu Roy \cite{basu} considers in detail the problem of finding the \emph{best coverage path} in the presence of transparent and opaque obstacles. The best coverage path (also called the \emph{Maximal Support Path}) is defined as the path between a source and a destination point which has the minimum \emph{cover value}, where cover value of a path is obtained by maximising over the distances of the points on the path from their corresponding nearest sensors. For the case of opaque obstacles, the crux of the algorithm presented is to compute the dual of the Constrained Voronoi Diagram (CVD) and to run a Bellman-Ford algorithm along the edges of the dual to find the shortest path. In the case of transparent obstacles, a visibility graph is used in place of the CVD.

Murray et al. \cite{p-centre-murray} extend the VDH-based solution \cite{das-06-BS_cover_Convex_area} to non-convex regions. The 1-centre problem, which is fundamental to solving the above problem, has been solved using efficient algorithms. The $p$-centre problem has the closest resemblance to the problem of total area coverage of a region $\mathcal{A}$. The $p$-centre problem is one of optimally placing the $p$ facilities (or sensors, equivalently) so that the distance of any point in $\mathcal{A}$ to its nearest sensor is minimised. It has both discrete and continuous versions. In the discrete version, the task is to select a subset of $p$ points from a large set such that placement of facilities on the points of the subset minimises the maximum distance of any other point in the set to its closest facility. The continuous version relaxes the constraint of having to place the facilities at the demand points. Instead, an underlying graph or map is defined and the facilities can be placed anywhere along the edges of the graph or the regions in the map so as to minimise the maximum distance of any demand point to its nearest facility.

The continuous $p$-centre problem may be formally stated as follows:
\begin{equation*}
\min_{(\hat{x}_j,\hat{y}_j)\in A,\ j=1,2\ldots,p}{\Biggl(\max_{(x,y)\in A}\biggl(\min_j{d_j(x,y)\biggr)\Biggr)}}
\end{equation*} where
\begin{tabbing}
$d_j(x,y)$ \quad\=: distance of point (x,y) to the $j^{th}$ facility\\
$(\hat{x}_j,\hat{y}_j)$ \>: coordinates of facility $j$\\
$A$ \>: the analysis region\\
$p$ \>: no. of facilities located\\
$(x,y)$ \>: reference point in $A$
\end{tabbing}
In terms of the $p$-centre problem, the problem of covering a plane region without obstacles reduces to
\begin{equation*}
\min_{C\le r_s} p
\end{equation*} 
where 
\begin{equation*}
C:=\min_{(\hat{x}_j,\hat{y}_j)\in A,\ j=1,2\ldots,p}{\Biggl(\max_{(x,y)\in A}\biggl(\min_j{d_j(x,y)\biggr)\Biggr)}}
\end{equation*} 
where $r_s$ is the sensing radius of a sensor.

The continuous $p$-centre problem is proven to be NP-hard \cite{results_Compl_pcenter_nimrod}. Yet the minimisation problem of the number of sensors with respect to the radius constraint has a surprisingly simple solution, as we will prove in the ensuing section.
\section{Coverage of a Plane Region sans Obstacles}\label{sect:plane}
\setcounter{figure}{0}
The problem of finding the minimum number of circles to cover an arbitrary two-dimensional region is well-studied and it is known that a tessellation of regular hexagons (inscribed in the coverage circle) solves the problem.  Kershner \cite{kershner39}, in 1939, gave the first formal proof of such a result, when the radius of the coverage circle reduces to zero for a fixed area to be covered. Particularly, he proved the following result.
\newtheorem{kershner}{Theorem}[section]
\begin{kershner}
Let $M$ denote a bounded and closed plane point set, $N(r)$ be the minimum number of circles of radius $r$ which can cover $M$ and $\Delta(M)$ represent the area of $M$. Then,
\begin{equation*}
\lim_{r\rightarrow 0}\pi r^2 N(r)=\frac{2\pi\sqrt3}{9}\Delta(M)
\end{equation*}
\end{kershner}
We note that the minimum number of disks of radius $r$ required to cover $M$ is $\frac{\Delta(M)}{\pi r^2}$. However a certain amount of overlap between neighbouring circles cannot be avoided, thus necessitating more disks than this minimum. Kershner's theorem actually says that the number of disks required is greater than this minimum by a factor $\frac{2\pi\sqrt3}{9} (=\frac{\pi r^2}{\frac{3\sqrt3}{2}r^2})$, which is the ratio of the area of a circle of radius $r$ to its inscribed hexagon. We extend this result to establish the optimality of hexagonal coverage of a rectangle using circles of a constant radius, as the area of the rectangle grows to $\infty$.

\subsection{Optimality of Hexagonal Tessellation for Infinite Plane Areas}
\newtheorem{circ-opt}[kershner]{Theorem}
\begin{circ-opt}
Let $l$ and $w$ be the length and width of a rectangular region $\mathcal{A}$ of area $A$, $N_{\Circle}^{r_s}(A)$ be the number of circles of radius $r_s$ that cover $\mathcal{A}$ and $A_{\hexagon}=\frac{3\sqrt3}{2}r_s^2$. Then,
\begin{equation*}
\lim_{\substack{l\rightarrow \infty\ \\w \rightarrow \infty}}\frac{N_{\Circle}^{r_s}(A)}{\frac{A}{A_{\hexagon}}}=1
\end{equation*}
\end{circ-opt}
\begin{proof}
The result follows from Lemmas 5 and 6 in \cite{kershner39}. 
From Lemma 5,
\begin{equation*}
\pi r_s^2 N_{\Circle}^{r_s}(A)>\frac{2\pi\sqrt 3}{9}\ (A-2\pi r_s^2)
\end{equation*}
\begin{equation*}
\frac{N_{\Circle}^{r_s}(A)}{\frac{A}{A_{\hexagon}}}>1-\frac{2\pi r_s^2}{A}
\end{equation*}
\begin{equation}
\liminf_{\substack{l\rightarrow \infty\ \\w \rightarrow \infty}}\frac{N_{\Circle}^{r_s}(A)}{\frac{A}{A_{\hexagon}}}\ge 1
\end{equation}
From Lemma 6, we have, for perimeter $p$ of $\mathcal{A}$,
\begin{equation*}
\pi r_s^2 N_{\Circle}^{r_s}(A)<\frac{2\pi\sqrt 3}{9}\ (A+2p r_s+16 r_s^2)
\end{equation*}
\begin{eqnarray*}
\frac{N_{\Circle}^{r_s}(A)}{\frac{A}{A_{\hexagon}}}<\frac{A+2p r_s+16 r_s^2}{A}\\
\frac{N_{\Circle}^{r_s}(A)}{\frac{A}{A_{\hexagon}}}<1+2r_s(\frac{1}{l}+\frac{1}{w})+\frac{16 r_s^2}{lw}
\end{eqnarray*}
\begin{equation*}
\limsup_{\substack{l\rightarrow \infty\ \\w \rightarrow \infty}}\frac{N_{\Circle}^{r_s}(A)}{\frac{A}{A_{\hexagon}}}\le 1
\end{equation*}
which completes the proof.
\end{proof}
Similar bounds can be established for coverage of $\mathcal{A}$ with hexagons.
\newtheorem{hex-opt}[kershner]{Theorem}
\begin{hex-opt}
Let $l$ and $w$ be the length and width of a rectangular region $\mathcal{A}$ of area $A$, $N_{\hexagon}^{r_s}(A)$ be the number of hexagons of circumradius $r_s$ that tessellate $\mathcal{A}$ and $A_{\hexagon}=\frac{3\sqrt3}{2}r_s^2$. Then,
\begin{equation*}
\lim_{\substack{l\rightarrow \infty\ \\w \rightarrow \infty}}\frac{N_{\hexagon}^{r_s}(A)}{\frac{A}{A_{\hexagon}}}=1
\end{equation*}
\end{hex-opt}
\begin{figure}[htbp]
\begin{center}
\includegraphics{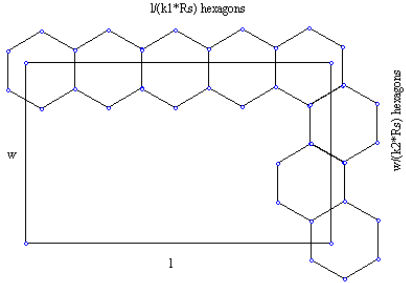}
\caption{Peripheral Hexagons displaying the finite edge effects which establish the lower bound for $\frac{N_{\hexagon}^{r_s}(A)}{\frac{A}{A_{\hexagon}}}$}
\end{center}
\end{figure}
\begin{proof}
The finite edge lengths dictate that not all hexagons employed of the tessellation occupy an area of $A_{\hexagon}$.
\begin{eqnarray*}
\frac{N_{\hexagon}^{r_s}(A)}{\frac{A}{A_{\hexagon}}}>1\\
\liminf_{\substack{l\rightarrow \infty\ \\w \rightarrow \infty}}\frac{N_{\hexagon}^{r_s}(A)}{\frac{A}{A_{\hexagon}}}\ge 1
\end{eqnarray*}
Let $k_1$ and $k_2$ be constants such that $\frac{l}{k_1r_s}$ and $\frac{w}{k_2r_s}$ refer to the number of hexagons along the length and width of the rectangle respectively. $k_1$ and $k_2$ are dependent upon the position and orientation of the tessellation used.
\begin{eqnarray*}
N_{\hexagon}^{r_s}(A)A_{\hexagon}-2A_{\hexagon}(\frac{l}{k_1r_s}+\frac{w}{k_2r_s})<A\\
\limsup_{\substack{l\rightarrow \infty\ \\w \rightarrow \infty}}\frac{N_{\hexagon}^{r_s}(A)}{\frac{A}{A_{\hexagon}}}\le 1
\end{eqnarray*}
\end{proof}
\newtheorem{opt-cor}[kershner]{Corollary}
\begin{opt-cor}
For a two-dimensional infinite plane region, tessellation with hexagons is optimal with regard to the number of hexagons used, i.e., 
\begin{eqnarray*}
\lim_{\substack{l\rightarrow \infty\ \\w \rightarrow \infty}}\frac{N^{r_s}_{\hexagon}}{N^{r_s}_{\Circle}}=1
\end{eqnarray*}
\end{opt-cor}
\section{Coverage in the Presence of \\Transparent Obstacles}\label{sect:trans_obs}
\renewcommand{\thefootnote}{\fnsymbol{footnote}}
In this section, we concentrate on extending Kershner's work to regions with transparent obstacles at arbitrary locations and of arbitrary shapes and sizes.
\subsection{Terminology}
Kershner's result establishes the fact that for an infinite plane area sans obstacles, tessellation with hexagons provides an asymptotic minimum with regard to the number of sensors used. Such a universal hexagonal tessellation (a hexagonal tessellation spanning the entire 2-D plane) is defined by two parameters: the centre of one of its hexagons and the orientation of the hexagons (i.e., say the angle made by any one of the diameters of the hexagon with the positive $x$-axis.) In this section, let $\mathcal{U}$ denote one such universal hexagonal tessellation of an infinite plane area. An area is said to be occupied by a \textit{transparent obstacle} if it is unavailable for placement of sensors, but is not impervious to the signal to be sensed. Let $\mathcal{A}$ be a finite and bounded plane area interspersed with transparent obstacles. Obstacle-free regions within $\mathcal{A}$ are referred to as `lands' and are \textit{accessible}, while those regions that are occupied by transparent obstacles are termed \textit{inaccessible}. An island is defined in the usual sense, i.e., an area of land surrounded by a transparent obstacle all around it. $\Delta(\mathcal{A})$ stands for the accessible area present within the region $\mathcal{A}$.

\subsection{Bounds on the Number of Sensors}
$\mathcal{A}$ is initially tessellated with hexagons from $\mathcal{U}$, and since $\mathcal{A}$ has transparent obstacles, some of the hexagons' centres may be found to lie in inaccessible regions. We call this initial tessellation the \emph{primary tessellation}. Some of the hexagons may lie entirely within the transparent obstacle and hence the corresponding sensor is not required to cover any land. Hexagons whose centres lie in a transparent obstacle, but which have some land requiring to be covered are termed \textit{anomalous hexagons}. Let $A$ denote the sum of areas of all \textit{normal hexagons} whose centres are accessible or `intact', and $A_o$ denote the sum of areas of all anomalous hexagons. The following theorem provides us with bounds on the number of sensors.

\begin{figure}[ht]
\begin{center}
\includegraphics[scale=.25]{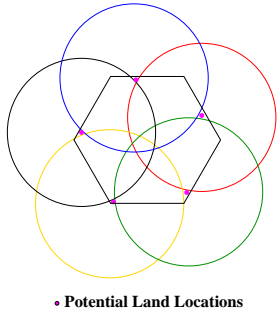}
\caption{Worst Case Positions of Islands within a Hexagon}
\label{fig:five-sensors_max}
\end{center}
\end{figure}

\newtheorem{bounds}{Theorem}[section]
\begin{bounds}\label{bounds}
For an area $\mathcal{A}$ with transparent obstacles, which has been tessellated with $\mathcal{U}$, the number of sensors, $N$, required for achieving total area coverage is bounded as follows:
\begin{equation}\label{eq:trans-obs-N-ub}
N\le \frac{A+5A_o}{A_{\hexagon}}
\end{equation}
\begin{equation}\label{trans-obs-N-lb}
N\ge \frac{A+A_o}{A_{\hexagon}}
\end{equation}
where
\begin{tabbing}
$A$\quad \=: sum of areas of all normal hexagons\\
$A_o$\ \>: sum of areas of all anomalous hexagons
\end{tabbing}
\end{bounds}
An intuition about the extreme situation can be gained from Figure~\ref{fig:five-sensors_max}. In the anomalous hexagon shown, we identify five five possible land locations and assume that the rest of the hexagon is occupied by the transparent obstacle. The circles shown are the coverage disks of the sensors placed at these locations. It could be seen that none of the five sensors aid in covering more than one land location, thereby making their usage indispensable.
We need the following lemmas in order to prove Theorem \ref{bounds}. 
\newtheorem{sixmaxlemma}[bounds]{Lemma}
\newtheorem{sixmaxlemmacor}[bounds]{Lemma}
\begin{sixmaxlemma}
No six points can be located in the closure of a regular hexagon of circumradius $r_s$ such that every pair of the points is at a distance $> r_s$ from one another.
\end{sixmaxlemma}
\begin{proof}
We prove the lemma by contradiction. Suppose there exist six such points $x_1 \ldots x_6$. They clearly belong to different sextants within the hexagon, as the maximum distance between two points in a sextant is $r_s$. Identify the sextants to which they belong. Drop a perpendicular from each of these points to the edge of the hexagon defining its corresponding sextant.

\textbf{Claim}: If we move the points $x_1 \ldots x_6$ along the perpendiculars to meet the edges at say, $x_1^\prime\ldots x_6^\prime$, distance between no pair of points decreases.
\begin{figure}[htbp]
\begin{center}
\includegraphics[scale=0.85]{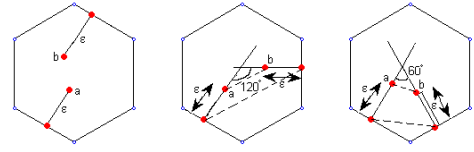}
\caption{Three cases for the points lying inside different sextants}
\end{center}
\end{figure}

\textbf{Proof of Claim}:\\
We identify three cases for the angles with which the perpendiculars meet:
\begin{itemize}
\item The perpendiculars are either parallel or meet at  $180^\circ$, in which case the claim is obvious.
\item The perpendiculars meet at  $120^\circ$. If $a$, $b$ are the initial distances of the points from the intersection of the perpendiculars and $\epsilon$ is the distance moved by the points along their respective perpendiculars ($\epsilon >0$), from cosine law, 
\begin{eqnarray*}
a^2+b^2+ab <(a+\epsilon)^2+(b+\epsilon)^2+(a+\epsilon)(b+\epsilon)
\end{eqnarray*}
\item The perpendiculars meet at  $60^\circ$. If the points, say $x_1$ and $x_2$ are at distances $a$ and $b$ respectively from the point of intersection, the distance between them is $a^2+b^2-ab$.  So, 
\begin{align*}
\epsilon^2+(a+b)\epsilon&\ge 0 \text{ } \Rightarrow\\
a^2+b^2-ab &\le(a+\epsilon)^2+(b+\epsilon)^2-(a+\epsilon)(b+\epsilon)
\end{align*}
\end{itemize}

Thus the claim is proved in each of the three cases and hence the six points could be safely assumed to be on the sides of the hexagon. Clearly, each of these points must lie on a different edge. Now join the six points to form a convex polygon. The polygon is convex since all interior angles are $< 180^\circ$.

\textbf{Claim}: 
If a convex polygon lies in the closure of another convex polygon, the inner one has a smaller perimeter than the outer one.

\textbf{Proof of Claim}:
Consider the six triangles, each of which shares a unique edge with the interior polygon, and has the other two edges along the perimeter of the outer polygon. Triangle law can be invoked in each of those triangles and the resultant inequalities can be added up to prove the claim. 

The perimeter of the assumed convex polygon is $>6r_s$, which is a contradiction to the previous claim. The existence of such a polygon is thus invalidated.
\end{proof}
\begin{sixmaxlemmacor}\label{lem:sixmaxlemmacor}
Five points can be located in the closure of a regular hexagon of circumradius $r_s$ such that no two of them are at a distance $\le r_s$ from each other.
\end{sixmaxlemmacor}
\begin{proof}
The proof follows from basic geometry and is omitted for the sake of brevity.
\end{proof}

\begin{proof}[Proof of Theorem \ref{bounds}]
The bounds can now be established using the previous Lemmas. We just need to observe that the hexagons under $A_o$ necessitate at most five sensors each. The land areas that make up $A$ require just one sensor per hexagon (Figure~\ref{fig:five-sensors_max}). Similarly, a lower bound on the number of sensors for covering any region arises in a land pattern where the anomalous hexagons demand no more than a single sensor.
\end{proof}

\subsection{1-Coverage of Anomalous Hexagons}
We propose an algorithm for coverage in the presence of transparent obstacles that involves recursively sliding the universal hexagonal tessellation $\mathcal{U}$ over the area that remains to be covered iteratively until no region is left uncovered. We recall the earlier notation of $A$ for sum of areas of hexagons with centre intact, i.e., normal hexagons and $A_o$ for sum of areas of hexagons with centre inaccessible but some land left to be covered, i.e., anomalous hexagons.

We define a \textit{cluster} as a group of anomalous hexagons contiguous to one another, where two hexagons are said to be contiguous if they share an edge. Let $P_1$, $P_2$, $\cdots P_k$ be the clusters formed among the $\frac{A_o}{A_{\hexagon}}$ anomalous hexagons. It could turn out that some (or even all) of the clusters may consist of a single hexagon. The goal now is to reduce the uncovered area in each of those clusters. The algorithm works separately on each cluster to reduce its uncovered area. Consider the following seemingly obvious fact about a hexagonal tessellation: When the centres of all the hexagons comprising a universal hexagonal tessellation are shifted by a specific distance and angle to points within their corresponding hexagons, the resulting points can be interpreted to be the centres of a new universal hexagonal tessellation. The common feature in all anomalous hexagons is the inaccessibility of their centres. This property enables us to find a suitable shift to their centres such that the maximum area is covered.

\newtheorem{periodicposn}[bounds]{Lemma}
\begin{periodicposn}\label{periodicposn}
Consider a hexagonal tessellation of $\mathbb{R}^2$. If the centres of each of the hexagons is shifted by a distance and angle to another point within the same hexagon, these shifted centres form a new hexagonal tessellation of $\mathbb{R}^2$.
\end{periodicposn}

\subsection{Analysis of the Algorithm}
The convergence of the algorithm can be established by noting that in every iteration, the superposed hexagonal tessellation will reduce some area that is yet to be covered, though perhaps not all. Let us denote by $L$ the set of all possible shifts of the centre of a hexagon to another point within the same hexagon. We note that $L$ is uncountably infinite and a shift is defined by the distance and angle with respect to a reference. Let $l \in L$ represent one such shift. Let $U_l:\mathbb{R}^+\rightarrow\mathbb{R}^+$ be a function that operates on a region with a specific area. If $\hat{\mathcal{A}}$ represents the uncovered region in a cluster of anomalous hexagons, $U_l(\Delta(\hat{\mathcal{A}}))$ denotes the area remaining to be covered in $\hat{\mathcal{A}}$ after the sensors at the centres of the anomalous hexagons in the cluster are shifted by $l$. Three properties of $U_l$ are identified:
\begin{enumerate}
\item For $\hat{\mathcal{A}}\subset\mathcal{A}$ and $\Delta(\hat{\mathcal{A}})\ne 0$, $U_l(\Delta(\hat{\mathcal{A}}))\le \Delta(\hat{\mathcal{A}})$ with strict inequality holding for at least one $l$.
\item $U_l(\Delta(\hat{\mathcal{A}}))=\Delta(\hat{\mathcal{A}}),\ \forall\text{ } l \Rightarrow \Delta(\hat{\mathcal{A}})=0.$
\item If $\hat{\mathcal{A}}\subset\mathcal{A}$, $\mathcal{A}^\prime\subset\mathcal{A}$ and $\hat{\mathcal{A}}\cap \mathcal{A}^\prime=\emptyset$, $U_l(\Delta(\hat{\mathcal{A}}\cup \mathcal{A}^\prime))=U_l(\Delta(\hat{\mathcal{A}}))+U_l(\Delta(\mathcal{A}^\prime))$
\end{enumerate}

The first property holds because as long as there is some uncovered region in any of the hexagons, the centres could be shifted to a point within them thus ensuring a strict reduction in uncovered area. The second statement follows as a corollary, because if no shifting of centres yields any reduction in uncovered area, it means that the entire region is occupied by a transparent obstacle.
The third and obvious fact justifies dealing with clusters of anomalous hexagons. The area reduction achieved by a particular shift of centres of geographically separated clusters of hexagons is equal to the sum of reductions achieved by shifting each of them separately. 

The algorithm works as follows. Every cluster of hexagons is considered separately. For every cluster, we choose a particular shift of the centres of hexagons from the set $L$ that leaves the minimum area uncovered. We understand that it is an infinite minimisation requiring approximation in order to be executed in real time. One iteration involves computing the optimal shifts for all clusters in the region and locating the sensors. Some of the anomalous hexagons may have been covered completely during the iteration. We neglect them and we identify the clusters among the remaining ones. We follow the same procedure until the entire region is covered. We now analytically prove that this algorithm stops after a finite number of iterations. 
\begin{figure}[htbp]
\centering
\subfigure[Initial location of obstacles]{
\includegraphics[scale=0.3]{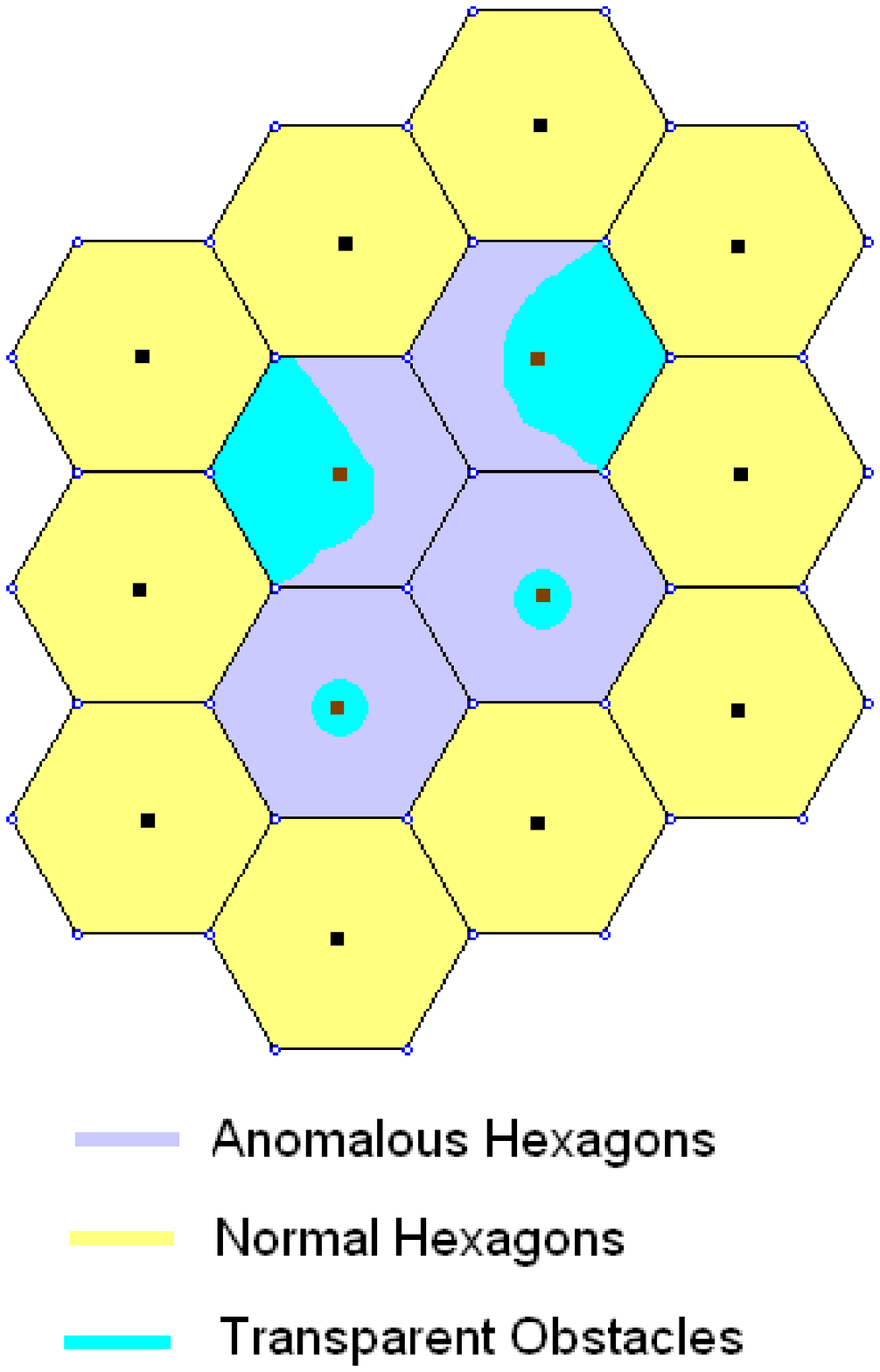}
\label{fig:iter0}
}
\subfigure[After an Iteration of the Algorithm]{
\includegraphics[scale=0.3]{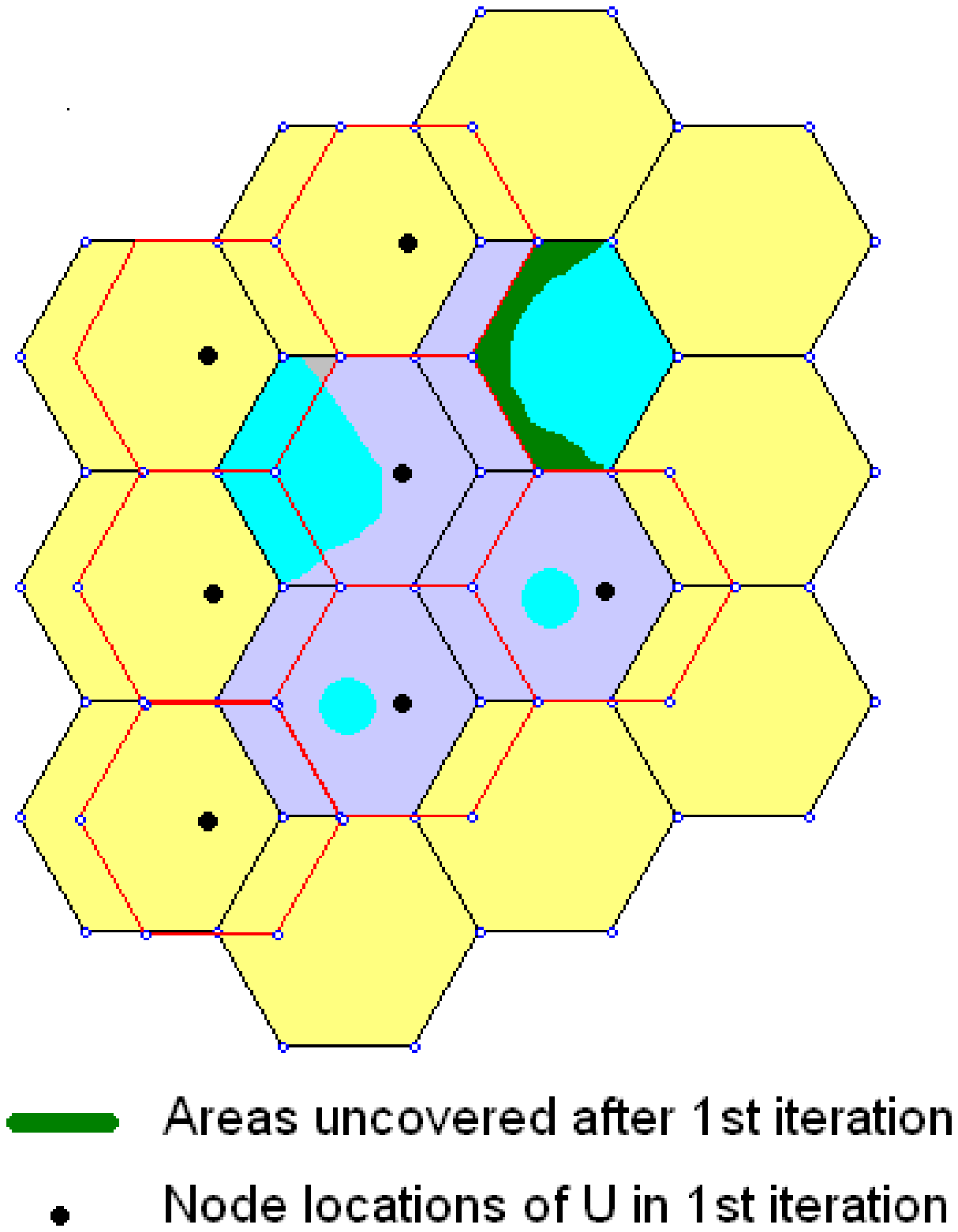}
\label{fig:iter1}
}\label{fig:Depiction of the Algorithm}
\caption{Depiction of the Algorithm}
\end{figure}

We define $\Delta(R^j)$ to be the area of the region remaining to be covered after the $j^{th}$ iteration and $\Delta(R_i^{j})$ as the area of the region that remains to be covered in the $i^{th}$ cluster after the $j^{th}$ iteration. Let $N^{(j)}$ be the number of clusters remaining after the $j^{th}$ iteration. We have, 
\begin{eqnarray*}
\Delta(R^j)=
\Delta(R^{j-1})-\Biggr[\sum_{i=1}^{N^{(j-1)}}\biggl(\Delta(R_i^{j-1})-\min_{l\in L}U_l(\Delta(R_i^{j-1})\biggr)\Biggr]
\end{eqnarray*}
where
\begin{align*}
\Delta(R^{j-1})&=\sum_{i=1}^{N^{(j-1)}}\Delta(R_i^{j-1})\\
\Delta(R^j) &=\sum_{i=1}^{N^{(j-1)}}\min_l U_l(\Delta(R_i^{j-1}))\\
&\le \min_l \sum_{i=1}^{N^{(j-1)}}U_l(\Delta(R_i^{j-1}))\\
&=\min_l U_l(\Delta(R^{j-1}))
\end{align*}
The penultimate equation tells us that the minimum uncovered area obtained by shifting the centres of anomalous hexagons spread over the entire cluster is greater than the minimum obtained by considering individual clusters. The last equation comes from the definition of a cluster and the third property. Therefore,
\begin{align}
\label{eqn:eq}
\Delta(R^j)&\le U_l(\Delta(R^{j-1}))\text{  }\forall\text{ } l\\
&< \Delta(R^{j-1})
\end{align}
\text{\ for at least one $l$, if }$\Delta(R^{j-1})\ne 0$. This follows from the first property. 
\newtheorem{conv-iterations}[bounds]{Lemma}
\begin{conv-iterations}\label{lem:conv-iterations}
The algorithm stops in a finite number of iterations if the number of anomalous hexagons is finite.
\end{conv-iterations}
\begin{proof}
The algorithm achieves strict reduction in area remaining to be covered in every iteration. This means that at least one sensor is placed in the course of an iteration. But the maximum number of sensors required in any area is bounded by Equation \ref{eq:trans-obs-N-ub}. Hence the number of iterations to convergence is also bounded by $\frac{5A_O}{A_{\hexagon}}$. 
\end{proof}
\newtheorem{conv}[bounds]{Lemma}
\begin{conv}
The sequence $\{\Delta(R^j)\}$ converges to zero.
\end{conv}
\begin{proof}
The sequence $\{\Delta(R^j)\}$ is a strictly decreasing sequence all of whose members are finite and positive. We use Lemma \ref{lem:conv-iterations} and the algorithm stops in $N$ iterations. Suppose say, for contradiction, the sequence has a positive limit $p>0$ and that further iterations of the algorithm do not cause a reduction in uncovered area.
\begin{eqnarray*}
\lim_{j\rightarrow N} \Delta(R^j)&=&p\\ 
\Delta (R^{N+1})&\le & U_l(\Delta (R^{N}))\text{  }\forall\text{ } l\\
\Rightarrow p &\le & U_l(p) \le p\\
U_l(p) &\le& p \text{ and } U_l(p) \ge p \text{ } \forall\text{ } l\\
U_l(p)&=&p\text{ } \forall\text{ } l \Rightarrow\text{ } p=0.
\end{eqnarray*}
The second equation follows from Equation \ref{eqn:eq}, whereas the third is a result of the first property of $U_l$. The last statement follows from the second property of $U_l$.
\end{proof}
\subsection{Pseudocode and Complexity}
In practice, the optimisation can be carried out over only a finite number of shifts. Hence we define a lattice of points inside a hexagon which represent the results of permissible shifts of the hexagon's centre. Thus we optimise over the set $L^\prime\subset L$. We begin with an initial hexagonal tessellation to cover the entire area inclusive of transparent obstacles. We then classify the hexagons as normal or anomalous depending on whether their centres are accessible or not.

Let $\mathcal{A}$ be uncovered region in a single cluster and $A$ be its area. Let $M$ be the number of clusters after the initial hexagonal placement and $|L^\prime|$ be the number of lattice points within each hexagon. Let $U_l(A)$ be the area remaining to be covered in a region with area $A$ after placement of a tessellation the centres of which are obtained by applying a shift $l\in L^\prime$ on the centres of the anomalous hexagons. Let us define $X$ to be the set of all coordinates of centres of hexagons when a shift $l\in L^\prime$ is applied to the existing centres. Let $x$ be an element of $X$ and $H(x)$ refer to the hexagon of circumradius $r_s$ centred at $x$. The following equation computes $U_l(A)$.
\begin{eqnarray}
U_l(A)=A-\Delta\Bigg[\bigg(\bigcup_X \big(H(x)\mathrm{I}_{\{x\in\mathcal{A}\}}\big)\bigg)\bigcap \mathcal{A}\Bigg]
\end{eqnarray}
In this equation, the second term on the RHS represents the area that gets covered because of the new tessellation. Some of the centres of the new tessellation could also be lying in the transparent obstacle and they may not contribute to coverage. The areas of those hexagons are excluded using the expression $\mathrm{I}_{\{x\in\mathcal{A}\}}$.  Also the areas of some of the new hexagons could partially overlap with regions that have already been covered. These are excluded from contributing to coverage again by means of the intersection with $\mathcal{A}$. The following pseudocode summarises the minimisation procedure adopted at every cluster. The result of the pseudocode, $best\_shift$ gives the particular shift of hexagons' centres that yields the maximum reduction in uncovered area in the cluster.\\
\textbf{\underline{Pseudocode}}:
\begin{tabbing}
$U_{best\_shift}(A)=\infty$\\
for $i$ = $1$ to $|L^\prime|$\\
    \quad\quad $best\_shift$=$\argmin(U_{best\_shift}(A),U_i(A))$\\
end
\end{tabbing}

The above procedure is repeated at each of the $M$ clusters present at the beginning of a particular iteration. If the number of anomalous hexagons in a cluster is given by $N_i$, $i=1,2\ldots M$, the complexity of the algorithm on a cluster $i$ is $O\big(N_i|L^\prime|^2\big)$. The overall complexity per iteration then becomes $O\big((\sum_{i=1}^{M}N_i) |L^\prime|^2\big)$, where $\sum_{i=1}^{M}N_i $ gives the total number of anomalous hexagons distributed over the $M$ clusters. The overall complexity of the procedure, given it stops in $N$ iterations, is then $O\big((\sum_{i=1}^{M}N_i) |L^\prime|^2N\big)$. Strictly speaking, the number of anomalous hexagons in every iteration reduces; yet the expression serves as an upper bound. If we take the number of permissible shifts of the centre of a hexagon to be a constant, the complexity becomes $O\big((\sum_{i=1}^{M}N_i )N\big)=O\big(\frac{A_o}{A_{\hexagon}}N\big)$, where $\frac{A_o}{A_{\hexagon}}$ gives the total number of anomalous hexagons present at the start of the procedure.
\section{Coverage in the presence of Opaque Obstacles}\label{sect:opq-obs}
A general version of problem of coverage in the presence of opaque obstacles has been studied previously in literature under the tag of art gallery algorithms. The na\"{i}ve art gallery problem is posed as follows:

\textit{ `If a location is said to be monitored by a particular guard if a direct line-of-sight path exists from the guard to the location, what is the minimum number of guards required to monitor a given polygon? What are their corresponding placements within the polygon?' }

The original art gallery problem dates back to 1973 and today the literature abounds with research articles on algorithms to solve the original problem and its myriad variants. The opaque obstacle problem can be mapped to one such variant - the art gallery problem with bounded visibility (i.e., since the sensor has a given coverage radius). The extra constraint of the guards having finite visibility is added to the above stated problem.

A polynomial time algorithm to this end was proposed in \cite{fast-posn-guards-kazazakis-02}. The algorithm essentially involves convexification of the input polygon and identifying a suitable location within each convex polygon to place a guard. The main downside of the algorithm is that some of the convex polygons tend to be extremely small and the utility of the guards present in such polygons is reduced. Moreover, the union of a few adjacent polygons may be monitored by a single guard in some circumstances, though the union may not necessarily remain convex. Bhattacharya et al. in \cite{exploring-bdd-visibility-amitavaetal-01} propose an incremental algorithm for coverage. The algorithm starts with an arbitrary point and computes the visibility polygon of the point. The next point is chosen to be on the boundary of the first point. The procedure is repeated until the entire region is covered. It is evident that this algorithm also trades off the number of sensors for guaranteed coverage.

\subsection{Problem Formulation}

We borrow two terms from the jargon of art gallery literature. A \textit{star polygon} is a polygon which can be monitored using a single guard, i.e., there exists at least one location inside the polygon such that a line-of-sight path can be constructed from that point to every other point in the polygon. Such a point is called a \textit{kernel}. It may be noted that all convex polygons are star polygons and all points in its interior are their kernels.
\begin{figure}[htbp]
\begin{center}
\includegraphics[scale=1.2]{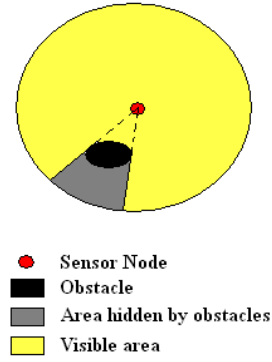}
\caption{RSP of a sensor in the presence of an obstacle}
\label{fig:opq_obs_RSP}
\end{center}
\end{figure}
Let $\mathcal{A}$ denote the region that requires sensor coverage. We define the \textit{`restricted star polygon (RSP)'} of a point $x$ as follows:
\begin{eqnarray*}
RSP(x)=\{y\ \in \mathcal{A}|\ \ \lVert y-x \rVert \le r_s,\  \lambda x+(1-\lambda)y\in\mathcal{A}\\ \ \forall\lambda \in [0,1] \}
\end{eqnarray*}
It is to be noted that the term `polygon' is a misnomer. The RSPs are circles for points on an infinite plane region without opaque obstacles. Figure~\ref{fig:opq_obs_RSP} depicts the RSP of a sensor in the presence of an opaque obstacle. In terms of the RSP, the problem at hand can be expressed as follows:
\begin{eqnarray}
\min_{\mathcal{A}\subseteq\cup_{i=1}^{N}RSP(x_i),\ \ x_1,x_2\ldots x_N\in \mathcal{A}} N
\end{eqnarray}
The goal as stated by the above formulation is to find the minimum set of points, the union of whose RSPs span the entire area $\mathcal{A}$.

\subsection{Complexity}
The general art gallery problem and many of its variants have been established to be NP-hard. In this section, we prove the NP-hardness of the stated problem.
\newtheorem{complexity}{Theorem}[section]
\begin{complexity}
The problem of total area coverage of a region with opaque obstacles is NP-hard.
\end{complexity}
\begin{proof}
The proof involves a mapping to the minimum set cover problem. The minimum set cover problem can be posed as below:

\textit{`Given a finite set $S$, a collection of subsets of $S$, namely $C$, and an integer $N$, does there exist a $C^\prime$ such that $C^\prime\subseteq C$, $\cup C^\prime=S$ and $|C^\prime|\le N$?'}

Let $\mathcal{A}$ be the region requiring coverage. To cast the decision version of the opaque obstacle problem as a min-set cover problem, we identify the following mappings:
\begin{eqnarray*}
\mathcal{A} &\Leftrightarrow& S\\
RSP(x),x\in \mathcal{A} &\Leftrightarrow& \text{sets constituting C}
\end{eqnarray*}
Are there $N$ points such that $\mathcal{A}\subseteq \cup_{i=1}^N RSP(x_i)? \Leftrightarrow$\\
$\quad\quad\quad$ Min Set-cover problem

The opaque obstacle problem is certainly in NP, as a polynomial-time algorithm could be devised to check if the RSPs in the solution combine to give $\mathcal{A}$. Hence the decision version of the problem is NP-complete and the optimisation problem is NP-hard.
\end{proof}

\subsection{Bounds on the Number of Sensors}

The bounds on the number of sensors required by a region containing opaque obstacles do not exist in the absence of some tailor-made assumptions. This can be intuited by observing that for every sensor placed inside the hexagon, the coverage zone of the sensor can be limited to an infinitesimal area around it by introducing an opaque obstacle that engulfs it on all directions.
Three assumptions are made to circumvent the above stated hitch in the derivation of bounds. Let $H_{R_O}(x)$ denote a hexagon centred at $x$ and of a given circumradius $R_O$, $R_O<<r_s$. The bounds on the number of sensors can be obtained in terms of $R_O$. Let $K$ represent the obstacle.

\begin{itemize}
\item $K$ is a convex set.

\item The obstacle must be large enough so that a hexagon of circumradius $R_O$ fits inside it. In other words, the obstacle $K$ is deemed \textit{valid} if and only if there exists at least one $x \in K$ such that $H_{R_O}(x) \subseteq K$.

\item None of the sensors employed should cater to the coverage needs of a region smaller than a hexagon of circumradius $R_O$. In other words, we do not care about the coverage of a region that does not contain at least $H_{R_O}$. Consider an anomalous hexagon where the entire region is occupied by opaque obstacles, except for a mass of land $A_m$. The assumption is that there exists at least one $x \in A_m$ such that $H_{R_O}(x) \subseteq A_m$.
\end{itemize}
The bounds can be derived in the light of the above assumptions. Let $\mathcal{A}$ be the region for which total area coverage is desired. We begin by tessellating $\mathcal{A}$ with an arbitrary hexagonal pattern $\mathcal{U}$. Complementary to the transparent obstacle case, we proceed to define \textit{`anomalous hexagons'}.
\newtheorem{ano-hex-def}[complexity]{Definition}
\begin{ano-hex-def}
If $x$ is the centre of one of the hexagons formed by the tessellation, the hexagon $H_{r_s}(x)$ centred at $x$ and of circumradius $r_s$ is said to be `anomalous' if either of the following conditions hold.
\begin{itemize}
\item $ x\in \mathcal{A}$ but $RSP(x)\cap H_{r_s}(x) \ne H_{r_s}(x) \cap \mathcal{A}$.
\item $x \notin \mathcal{A}$ and $\exists$ at least one $y$, $y\in H_{r_s}(x)$ and $y\in\mathcal{A}$.
\end{itemize}
\end{ano-hex-def}

\begin{figure}[htbp]
\begin{center}
\includegraphics[scale=1]{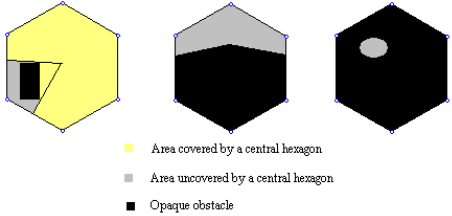}
\caption{Anomalous hexagons}
\label{fig:ano_opq_obstacles}
\end{center}
\end{figure}
The first proviso is applicable whenever we have a hexagon whose centre is on land, but has some opaque obstacle lying elsewhere within it. The second condition applies to hexagons whose centres are on obstacles, but there is some region uncovered within the hexagon. The area of such regions are lower bounded by our assumptions. Figure~\ref{fig:ano_opq_obstacles} presents three different instances of anomalous hexagons, the first of which falls under the first alternative and the rest under the second alternative of the definition.

\newtheorem{opq-bounds}[complexity]{Theorem}
\begin{opq-bounds}
For a region $\mathcal{A}$ with opaque obstacles, which has been tessellated with $\mathcal{U}$, the number of sensors, $N$, for total area coverage is bounded as follows:
\begin{equation*}\label{Nopq-ub}
N\le \frac{A+\frac{n}{3}A_o}{A_{\hexagon}}
\end{equation*}
where $n$ is the largest number of hexagons of circumradius $R_O$ that could be present (partially or completely) inside a hexagon of circumradius $r_s$.
\begin{equation*}\label{Nopq-lb}
N\ge \frac{A+A_o}{A_{\hexagon}}
\end{equation*}
\end{opq-bounds}

\begin{proof}
Consider a single anomalous hexagon. We need to compute the maximum number of sensors that may be required by this hexagon. We create a tessellation of hexagons of circumradius $R_O$ within the hexagon. Let the total number of such hexagons be $n$. The maximum number of sensors necessitated by the $r_s$-hexagon can be computed by identifying the corresponding distribution of obstacles and plane areas within the hexagon. The assumptions that we have made ensure that the smallest area that a sensor may be required to cover is equal to the area of a hexagon of circumradius $R_O$. This happens when one of the hexagons in the tessellation has uncovered area, but all of its neighbours consist of only obstacles. The maximum number of sensors are demanded by a region whose uncovered area appears as hexagons of circumradius $R_O$ and are duly covered by obstacles all around.

Thus in the tessellation, we segregate the $n$ hexagons into $N$ classes such that hexagons of a particular class are plane uncovered areas and the ones of other $N-1$ classes are occupied by opaque obstacles. We need a class arrangement such that no two plane hexagons are adjacent to each other, since a single sensor would be sufficient to cover both in that case. The problem is the same as the frequency allocation problem to a location partitioned into hexagonal cells where we ensure that adjacent cells do not share the same frequency. From standard results on the problem, we know that $N=i^2+ij+j^2$, $i,\ j \in \mathbbm{Z}$ where $i$ and $j$ are the number of hexagons traversed along two axes separated by $120^\circ$, to reach a cell of the same frequency as the starting cell. We also know that the densest packing of cochannel cells occurs for $N=3$.
\begin{align*}
&\text{Maximum number of sensors required}\\
&=\max_{N=i^2+ij+j^2,\ i,\ j \in \mathbbm{Z}} \frac{n}{N}\\
&=\frac{n}{3}
\end{align*}
The lower bound is attained if each of the anomalous hexagon requires no more than a single sensor for its coverage.
\end{proof}
\subsection{Algorithm and Convergence}
\begin{figure}[htbp]
\begin{center}
\includegraphics[scale=0.75]{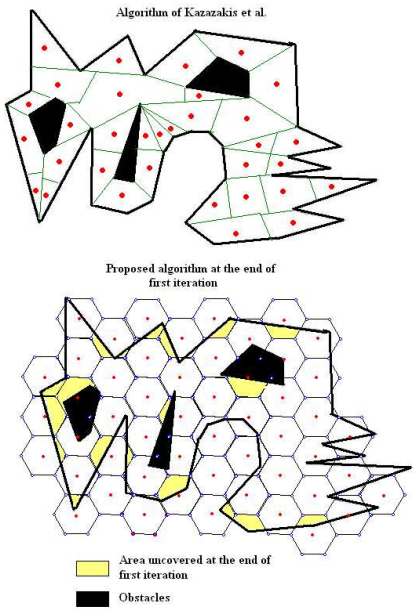}
\caption{A comparison of Algorithms}
\label{fig:opq_comp}
\end{center}
\end{figure}
The algorithm proposed for the opaque obstacle problem has the same ingredients as the one stated for coverage with transparent obstacles.  However, some extra computation arises due to the need to compute the RSP of each point involved in a particular placement.
As before, the entire area is tessellated with hexagons and the anomalous hexagons are identified. A \textit{`cluster'} is defined as a group of anomalous hexagons contiguous to one another. In a particular cluster, let $\mathcal{A}$ be the area interspersed with opaque obstacles and requiring coverage and $A$ be its area. Let $M$ be the number of clusters after the initial hexagonal placement and $N$ be the number of lattice points within each hexagon. Let $U_l(A)$ be the area remaining to be covered in a region with area $A$ after placement of a tessellation whose centres are moved in accordance with the shift $l \in L^\prime$. Let us define $X$ to be the set of all coordinates of centres of hexagons which result from the application of the shift $l$ on the primary tessellation. Let $x$ be an element of $X$ and $H_{r_s}(x)$ refers to the hexagon of circumradius $r_s$ centred at $x$.
\begin{eqnarray}\label{eqn:Ul-opq}
U_l(A)=A-\Delta\Bigg[\bigg(\bigcup_X \big(H(x)\mathrm{I}_{x\in\mathcal{A}}\big)\bigcap RSP(x)\bigg)\bigcap \mathcal{A}\Bigg]
\end{eqnarray}
The equation is nearly the same as for transparent obstacles, except for the RSP factor coming into play. The factor accounts for the fact that the area that gets covered in a particular iteration must lie within the RSP of the sensor, besides satisfying the usual criteria. 
Figure~\ref{fig:opq_comp} compares the performance of the proposed algorithm with an identical input polygon in \cite{fast-posn-guards-kazazakis-02}. The superiority of the proposed algorithm is clearly evident. The convergence results derived in the previous section also hold for this algorithm, with the sole difference lying in the definition of $U_l$ in Equation \ref{eqn:Ul-opq}.
\section{Conclusion and Future Work}\label{sect:concl_future-work}
\begin{figure}[htbp]
\centering
\subfigure[2-coverage]{
\includegraphics[scale=0.3]{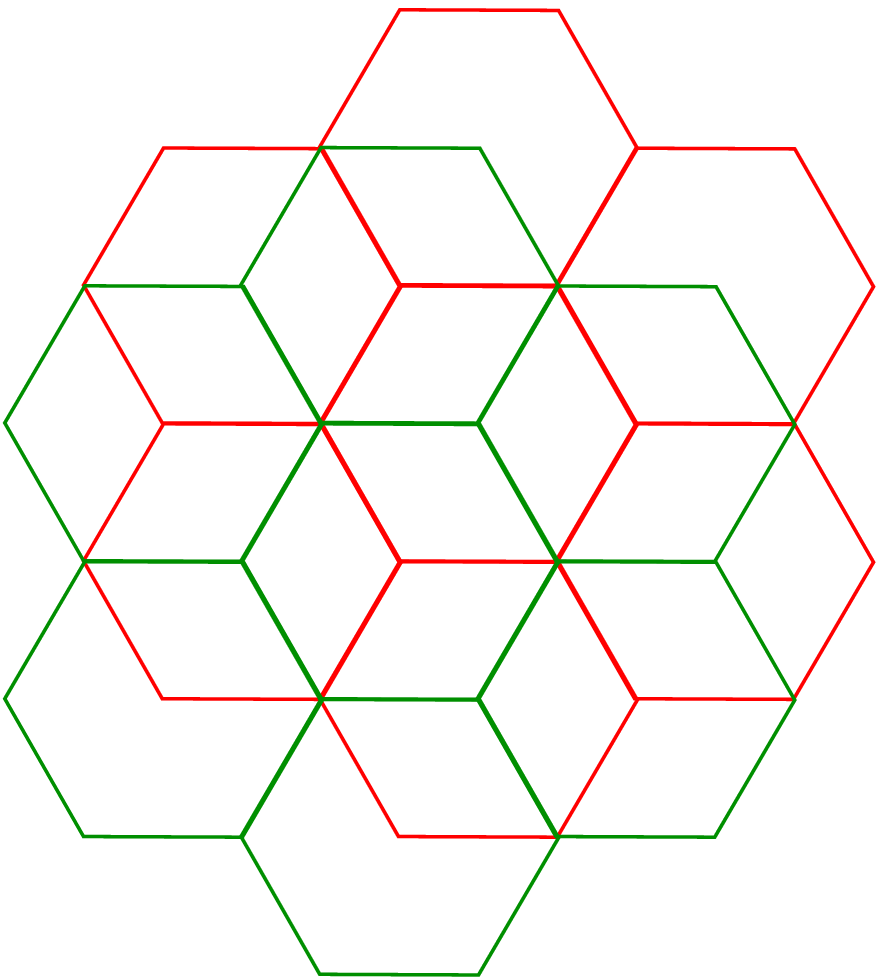}
\label{fig:2cov}}
\subfigure[3-coverage]{
\includegraphics[scale=0.3]{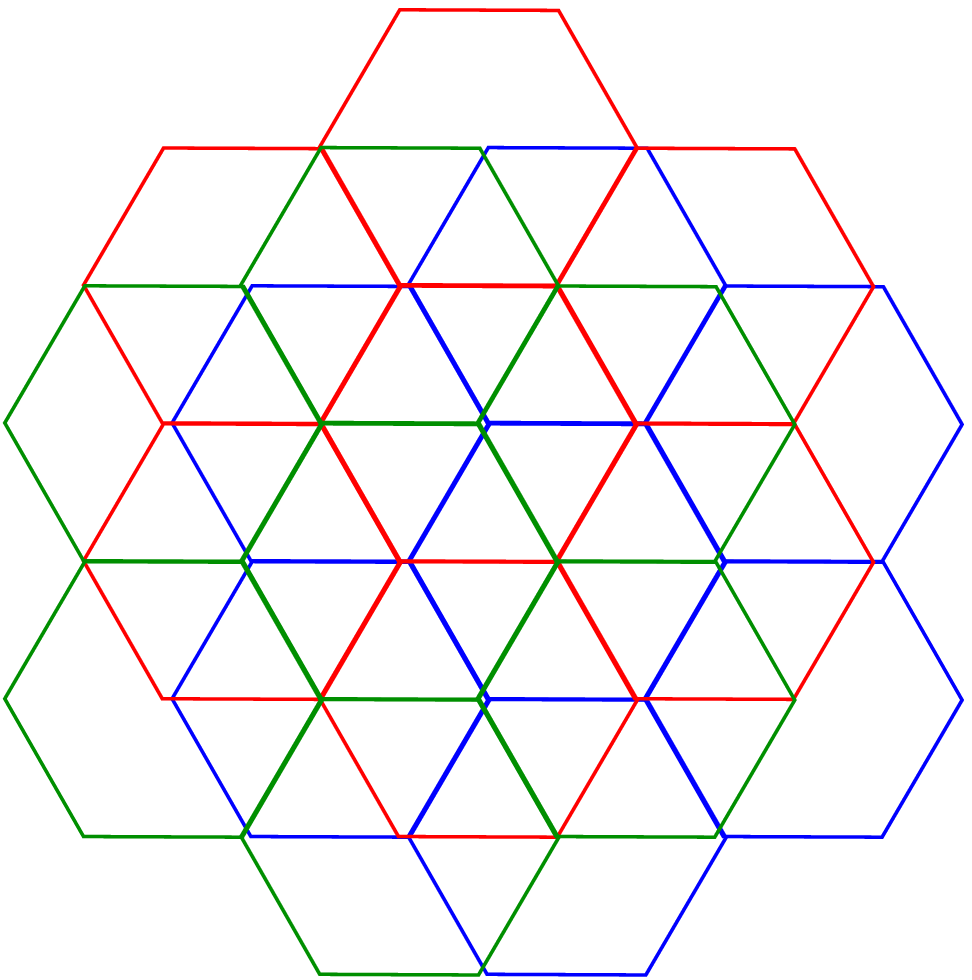}
\label{fig:3cov}}
\caption{Standard Procedure for 2 and 3-coverage of a plane region sans obstacles}
\label{fig:standard-2-3-cov}
\end{figure}

The results presented in this paper also lend themselves to solving the problem of $k$-coverage, for $k$=2, 3. Figure~\ref{fig:standard-2-3-cov} depicts the standard shift-based procedure for achieving 2 and 3-coverage. It is to be noted that the procedure ensures that the sensors of different tessellations are spaced sufficiently apart.

2 and 3-coverage can be achieved in the presence of obstacles by ensuring that the shifted tessellations are all 1-covered, which is in turn accomplished by the application of the proposed algorithm separately on each of the shifted tessellations. A recent work \cite{chai-dinesh08} gives asymptotically optimal solutions for $k$-coverage problems of arbitrary $k$ for plane areas. A possible future extension of the work of this paper is to obtain optimal $k$-coverage placements for arbitrary $k$ in the presence of transparent and opaque obstacles.

\end{document}